# Plasmon Enhanced Solar-to-Fuel Energy Conversion

I. Thomann[1]*, B.A. Pinaud[2], Z. Chen[2], B.M. Clemens[1], T.F. Jaramillo[2], Mark. L. Brongersma[1]*

[1]Geballe Laboratory for Advanced Materials, 476 Lomita Mall, Stanford, California 94305-4045

[2]Department of Chemical Engineering, 381 North-South Mall, Stanford University, Stanford, California 94305-5025

* To whom correspondence should be addressed.

E-mails: ithomann@stanford.edu, brongersma@stanford.edu.

**Future generations of photoelectrodes for solar fuel generation must employ inexpensive, earth-abundant absorber materials in order to provide a large-scale source of clean energy. These materials tend to have poor electrical transport properties and exhibit carrier diffusion lengths which are significantly shorter than the absorption depth of light. As a result, many photo-excited carriers are generated too far from a reactive surface, and recombine instead of participating in solar-to-fuel-conversion. We demonstrate that plasmonic resonances in metallic nanostructures and multi-layer interference effects can be engineered to strongly concentrate sunlight close to the electrode/liquid interface, precisely where the relevant reactions take place. By comparing spectral features in the enhanced photocurrent spectra to full-field electromagnetic simulations, the contribution of surface plasmon excitations is verified. These results open the door to the optimization of a wide variety of photochemical processes by leveraging the rapid advances in the field of plasmonics.**

Solar fuel generation based on inexpensive, earth-abundant materials constitutes one potentially viable option to satisfy the demand for a terawatt scale renewable source of energy that can be stored and used on demand[1]. The efficiency of solar water splitting[2,3] based on earth-abundant materials made using scalable processing techniques has remained low despite intensive research efforts since the 1970s. One of the underlying reasons for the observed inefficiency is that many of these materials exhibit a large mismatch between the length scales over which photon absorption takes place (up to micrometers), and the relatively short distances over which electronic carriers can be extracted (often limited to a few 10's of nanometers). One possible approach to circumvent this challenge is to synthesize nanostructured electrodes in which the photon propagation and charge transport directions are orthogonalized. This type of geometry can be accomplished in wire arrays[4-6] or other nanostructures with large surface-to-volume ratios[7]. We pursue a new approach that is aimed at the use of metallic nanostructures for better photon management. The basic concept is illustrated in Fig. 1a, which shows how a metallic nanoparticle can concentrate light near a semiconductor/liquid junction to produce more photo-carriers that can reach the interface and participate in desired reactions.

Recently, the use of metallic nanostructures has proven extremely effective in enhancing the efficiency of thin film solar cells whose performance is constrained by similar issues [8-14]. Metallic nanoparticles support collective electron oscillations, known as surface plasmons (SPs). At specific frequencies these charge oscillations can be driven resonantly and produce intense light fields near the metal and cause strong light scattering. As such, metal particles serve as optical antennas that operate analogously to larger radio antennas[15]. The light concentration and scattering effects were effectively harnessed to trap light inside the absorbing semiconductor layers of a solar cell and to ultimately enhance the power conversion efficiency. The SP resonance frequency of a metal particle can be tuned across the ultra-violet, visible and near-infrared parts of the electromagnetic spectrum through a choice of its size, shape, and dielectric environment[16]. This notion has enabled broadband enhancements of the photocurrent across the wide solar spectrum. Rapid developments in this area were founded upon initial, convincing demonstrations that plasmonic effects are capable of boosting efficiency[8]. Key to demonstrating the importance of plasmonics in enhancing solar cell efficiency was the presence of characteristic plasmonic features in the photocurrent spectra from cells with metallic nanostructures. In this paper we aim to leverage the existing knowledge base on plasmon-enhanced photovoltaics to convincingly demonstrate that plasmonic effects can also enhance solar energy conversion to fuels.

The enhancement of solar energy conversion to fuels imposes different requirements on the plasmonic nanostructures than solar cells. This fact can be appreciated by analyzing the materials properties of a typical photoelectrode and the band diagram of a standard photoelectrolysis cell used for water splitting. In this paper we describe a specific cell that employs iron oxide. This n-type semiconductor can be viewed as a prototypical system that is of topological interest and which shares many features with other candidate materials for future large-scale solar fuel production. Iron oxide ($\alpha$-$Fe_2O_3$; hematite) is a material that has been extensively researched for water splitting, mainly because it is corrosion-stable, inexpensive, and earth-abundant. Its band gap is around 2.1 eV, close to an ideal value for water splitting by a single semiconductor material. One major shortcoming of iron oxide as a photoelectrode material is its poor electronic transport, with a minority carrier diffusion length on the order of 2-4 nm[17] or 20 nm[18]. In addition, $Fe_2O_3$ is a relatively weak absorber in the 500-600 nm range (~ 0.1 – 1 µm absorption length, far longer than the generated photocarriers can travel).

Figure 1b shows the band diagram of a cell which employs iron oxide as the photoanode. Upon illumination, photogenerated holes move toward the semiconductor/liquid interface and oxidize water to produce oxygen. The electrons generated in this oxidative process are transported through an external circuit to a metal counter electrode where they can drive hydrogen evolution. Using this circuit, we can measure the wavelength-dependent photocurrent to determine the photocurrent spectra. Driving the water splitting reaction with iron oxide requires an applied potential due to its mismatched conduction band alignment[19]. This limitation can be overcome by using the material in a tandem cell configuration with two semiconductor electrodes[20]. The band diagram suggests that issues with poor charge transport might be alleviated if metallic nanoparticles could be used to concentrate incident sunlight as close as possible to the semiconductor/liquid interface. Here, the space charge layer can quickly separate the photocarriers and deliver the holes to the $Fe_2O_3$/$H_2O$ interface. For stable operation, the metallic nanoparticles will need to be chemically stable as well.

In the following, we describe a set of experiments that show that (1) metallic nanostructures can enhance photocurrent in spectral regions near the surface plasmon resonance and that (2) the spectral dependence of the photocurrent spectra is characteristic of plasmonic structures. Our aim is to separate plasmonic effects from other chemical/physical effects that can impact the photocurrent. To this end, we show that the use of different types of metallic nanostructures and the different placement of such particles with respect to the photoelectrode material affect the photocurrent spectra in a predictable way which is consistent with plasmonic effects.

In our first experiment, we aimed to demonstrate that metal nanoparticles can enhance photocurrents while avoiding catalytic effects. To this end, we utilized Au nanoparticles coated with a non-reactive silica ($SiO_2$) shell. Empirically, we found the strongest photocurrent enhancements with 50 nm diameter gold particles coated with a thin silica-shell. Figures 2b,d show SEM images of two different $Fe_2O_3$ samples containing these particles. Figure 2b shows a sample with these nanoparticles deposited onto a transparent conductive oxide electrode that was covered by a 100 nm thick iron oxide film. The presence of the nanoparticles can still be discerned underneath the thin $Fe_2O_3$ film. In Fig. 2d identical core-shell nanoparticles were deposited on top of a 90 nm thick iron oxide film. The particles are clearly visible and are exposed to the aqueous solution during the electrochemical experiments.

Figures 2a,c show spectral photocurrent measurements on these electrodes taken in a three-electrode photoelectrochemical cell (see Methods). The photocurrent enhancement $\varepsilon$ spectra were generated by normalizing the photocurrent $j_{NP}(\lambda)$ from a region with particles to the photocurrent $j_{Ref}(\lambda)$ obtained in a region without nanoparticles (i.e. $\varepsilon = j_{NP}(\lambda)/ j_{Ref}(\lambda)$). In Fig. 2a a strong photocurrent enhancement is observed at wavelengths longer than 550 nm, whereas the photocurrent enhancement is close to unity at short wavelengths. The strongest resonant peak enhancement of 11 x is observed at 610 nm. Under AM1.5 illumination, plasmonic effects increase the wavelength-integrated photocurrent by 7 % over the planar reference structure (see Supplemental Fig. S8).

In Fig. 2a,c we also show simulated absorption enhancement spectra obtained using full-field electromagnetic simulations (see Methods). In these simulations, the absorption enhancement was determined by calculating the dissipated power in a small iron oxide region near a metal nanoparticle and near the liquid interface, and taking a ratio of this dissipated power in a structure with the plasmonic nanoparticle present and in a planar reference structure. It is expected that the measured spectral photocurrent scales linearly with the dissipated power in the probe volume. There is an excellent qualitative agreement between the simulations and the experimental results – the matching spectral shapes reveal that plasmonic effects are what give rise to improved photoactivity. An exact quantitative comparison of the power dissipation and the photocurrent is however complicated by a variety of factors. Among the most important are that typical, high performance samples (1) possess substrate and iron oxide film roughness, (2) exhibit carrier diffusion lengths that are sample and potentially position dependent, (3) feature a non-uniformity in the nanoparticle size, shape, and density, and (4) could show positive catalytic effects[21] and enhanced charge separation or recombination caused by the introduction of metallic nanostructures[22-24] that can add to the complexity of the system. Our arguments are thus primarily based on the spectral dependence of the photocurrent enhancement, which we have found to be robust against the effects described above, enabling us to confirm the plasmonic origin of the observed enhancements.

The blue line in Fig. 2a shows the simulated absorption enhancements with a single broad peak that closely resembles the spectral shape of the experimentally observed enhancements. The exact peak position is very sensitive to the dielectric environment of the metallic particle. The resonance position is controlled by optical properties of the material surrounding the particle[25]. It is known that even nanometer scale variations in the thickness of a silica shell on a Au particle can cause observable shifts in the resonance position[26]. Based on the position of the nanoparticles within the $Fe_2O_3$ film, it is expected that the resonant wavelength of the Au particles lies between calculated values for Au particles surrounded by $SiO_2$ and Au particles surrounded by $Fe_2O_3$. Our simulations reveal that the SP resonance of 50 nm gold particles surrounded by a thin $SiO_2$ shell is around 550 nm and that the resonance moves to 630 nm when this shell is replaced by $Fe_2O_3$ (Supplemental Figure S1). The experimentally observed peak falls squarely between those values. As the exact morphology and composition of the shells are unknown after the sample fabrication, we followed two approaches to mimic a mixed $SiO_2/Fe_2O_3$ environment in our simulations. Figure 2a shows simulation results with shells composed of a mixed $SiO_2$ - $Fe_2O_3$ medium with equal amounts of each component. Similarly good agreement with the experimental results was obtained in simulations using thinner silica shells rather than mixed shells.

Figure 2c shows photocurrent enhancements for a different configuration with the silica-coated Au particles deposited on top of iron oxide. For this electrode the measured photocurrent enhancement spectrum (red symbols) again displays a strong resonant peak around 590 nm, with a ~11 x peak enhancement. It also shows an almost spectrally flat region with an ~ 8 x enhancement. The simulation (blue line) is consistent with a plasmonic origin for the peak, but plasmonic effects cannot account for the wavelength-independent background enhancement (the electromagnetic simulation shows an enhancement close to unity at 500 nm). We speculate that this wavelength-independent enhancement in the presence of silica shells at the semiconductor/liquid interface may arise from morphological changes of the iron oxide in the presence of the nanoparticles[27], from additional n-doping of the iron oxide surrounding the particles[27, 28], or from possible catalytic effects on water oxidation by the particles' silica-shells in contact with the liquid. The additional spectral oscillations in Fig. 2c compared to Fig. 2a arise from multilayer inferences with a thicker ~940 nm FTO (fluorine-doped tin oxide) substrate employed here, compared to the ~150 nm ITO (tin-doped indium oxide) substrate employed in Figure 2a.

The measurements on the silica-coated Au particles show significant enhancements and reasonable agreement with a simple plasmonic model. The coatings on these particles give rise to several added benefits. They can be deposited at a high surface coverage without significant red-shifts of the plasmon resonance due to particle-interactions (see refs[29, 30] and Supplementary Information Fig. S2). The high photocurrents also show that the shells are effective in preventing severe nanoparticle-induced charge recombination, which represents a well known problem for the use of metallic nanoparticles[24]. Despite the many benefits of these coated particles, they do not serve as an ideal optical model system as their shape tends to be non-spherical, the silica shell is non-dense[31, 32] and thermal processing may cause further shell deformation and interdiffusion of atomic species[33]. For this reason, we also explored the simplest possible model system which is a bare, spherical Au nanoparticle.

Figure 3 shows the measured (black symbols) and simulated (blue lines) photocurrent enhancement spectra for bare, 50 nm diameter, Au particles on top (Fig. 3a) and at the bottom (Fig. 3c) of an iron oxide film. A comparison of the two spectra shows that the peak position and lineshape of photocurrent enhancements depend critically on the position of the nanoparticles in the iron oxide film. Particles on top of the absorber film (Fig. 3a,b) produce an asymmetric lineshape of the photocurrent enhancement, while particles on the bottom (Fig. 3c,d) produce a more symmetric spectral feature. Both effects are true signatures of plasmonic behavior, as will be explained below. Our simulations (Fig. 3b) and experiments (Supplementary Information Fig. S3) show that these spectral features persist for all investigated film thicknesses. These types of spectral features have also been observed in plasmon-enhanced solar cells[9] and played a critical role in proving the relevance of plasmonic effects in enhancing photovoltaics.

Figures 3b,d show the dependence of the simulated absorption enhancements in the $Fe_2O_3$ with $Fe_2O_3$ film thickness. The plasmon resonance found in simulations of 50 nm gold particles on top of (embedded in) semi-infinite $Fe_2O_3$ films is indicated by the white dashed lines. For particles on top of relatively thin films we find the maximum light concentration and absorption enhancements at slightly longer wavelengths than the particles' surface plasmon wavelength. Furthermore, the wavelength of maximum enhancement shifts linearly with the thickness of the iron oxide film, consistent with a simple multilayer interference effect. From these observations it can be concluded that the final absorption enhancement depends on an interplay between plasmonic and multilayer interference effects.

The mechanism by which plasmonic resonances allow one to control the flow of electromagnetic energy into the absorber can be understood from plots of the electric field distributions in our structures. Figure 4a shows two simulation geometries consisting of a planar reference structure (left) and the same structure with a Au nanoparticle placed on top of the $Fe_2O_3$. This configuration results in the asymmetric photocurrent spectrum. The probe region (PR) in which absorption enhancements were calculated is indicated in black. In Fig. 4b the simulated absorption enhancements in the probe regions are shown together with the dissipated power in the nanoparticle. The spectra are complicated by multilayer interference effects that produce oscillations in the spectra in addition to the strong plasmonic resonance peaks (see the dashed line in Fig. 4b for a comparison to the plasmonic peak shape of a semi-infinite iron oxide film). These interferences can be used to one's advantage as they enable additional spectral tuning and larger enhancements (see Fig. 3b,d).

The simulated electric field distributions generated by bare Au nanoparticles on top of iron oxide can explain the asymmetric lineshape versus wavelength. We will show that it results from the interference between the incident light wave and the scattered fields from the particle. Figures 4c-e show these distributions for wavelengths of 500 nm (shorter than the plasmon resonance wavelength, $\lambda < \lambda_{SP}$), 590 nm ($\lambda = \lambda_{SP}$), and 650 nm ($\lambda > \lambda_{SP}$) respectively. We plot the x-component of the electric field without the nanoparticle present, the x-component of the scattered- and near-field of the nanoparticle, and the modulus of the total field in the presence of the nanoparticle. At wavelengths shorter than the SP resonance wavelength, the electron oscillations in the metal particle can no longer follow the rapid field variations of the incident light wave and the scattered fields pick up a significant phase shift. This leads to a destructive interference between the incident light and the scattered light in the forward direction (i.e. below the particle and in the probe region). This resulting lower light intensity is the cause of a lower photocurrent at wavelengths where $\lambda < \lambda_{SP}$ (black region in Fig. 4c). In contrast, Fig. 4e shows that at longer wavelengths ($\lambda > \lambda_{SP}$), the interference in the forward direction is constructive, effectively increasing the light intensity in the probe region in the absorber film and enhancing the photocurrent. We note that interference effects also govern the observed enhancements for Au particles placed at the bottom of an $Fe_2O_3$ film (Fig. 3c,d). Here, constructive interference in the backward direction enables strong enhancements at wavelengths shorter than the SP resonance (500 - 600 nm region for the thinnest films in Fig. 3d). This data not only highlights the importance of plasmonic effects in the generated photocurrent, it also provides practical considerations for the design of future plasmon-enhanced photoelectrodes.

We have shown that plasmonic structures can circumvent the traditional compromise between photon absorption and carrier extraction in photoelectrodes made from inexpensive, earth-abundant semiconductor materials. Future efforts should be aimed at designing plasmonic enhancement structures that are effective at shorter wavelengths. Gold is limited in this respect, as its plasmon resonance is around 530 nm or longer, depending on the host medium. Metals with plasmon resonances in the near-UV, such as aluminum or silver, can be tuned into the region of interest by choice of size, shape, dielectric environment, and multilayer interference effects. Such metals need to be protected against corrosion, e.g. by covering them with inert shells, and we hope that our work will stimulate further materials research efforts in this direction.

We anticipate that the concepts described here will also be highly relevant to the development of future, more efficient multi-junction photoelectrochemical cells, where sunlight is split into multiple spectral components, each of which requires its own optical tailoring and enhancement strategies. Finally, it is worth noting that plasmon-enhanced photon management will be of value to other high-impact photocatalysis applications, including the removal of toxic compounds from the environment and large-scale water purification[34].

We acknowledge fruitful collaborations and discussions with Stacey Bent's and Bruce Clemens' group, and thank Jonathan Bakke for ALD work. We gratefully acknowledge Wenshan Cai for help with simulations and Tom Carver for e-beam evaporations. This work was supported by the Center on Nanostructuring for Efficient Energy Conversion (CNEEC) at Stanford University, an Energy Frontier Research Center funded by the U.S. Department of Energy, Office of Science, Office of Basic Energy Sciences under Award Number DE-SC0001060. I. Thomann and M.L. Brongersma also acknowledge support from Samsung. I. Thomann gratefully acknowledges a postdoctoral fellowship from the Deutsche Forschungsgemeinschaft (DFG). B. Pinaud gratefully acknowledges a graduate fellowship from the Natural Sciences and Engineering Research Council of Canada.

**Author contributions:**
I.T., M.L.B., T.F.J., and B.C. conceived the experiments. I.T. fabricated the nanostructures. I.T. and B.P. carried out the photo-electrochemical experiments. I.T. carried out the electromagnetic simulations. I.T. and M.L.B. analyzed the data. Z.C. contributed Fig. 1b. All authors were involved in interpreting the data and writing the manuscript.

**Methods:**
Preparation of iron oxide photoelectrodes: Iron oxide ($\alpha$-Fe$_2$O$_3$; hematite) was grown by e-beam evaporation of iron onto TCO (transparent conducting oxide) substrates, followed by calcination in air at 600 deg Celsius. X-ray diffraction analysis (see Supplementary Information) demonstrated that hematite films ($\alpha$-Fe$_2$O$_3$) were grown. Samples with nanoparticles at the bottom or on the top of the iron oxide film were prepared as follows: For samples with nanoparticles deposited at the bottom, the nanoparticles were drop-casted on the TCO substrate before the iron was evaporated and calcined. For samples with nanoparticles on the top, the nanoparticles were drop-casted on top of the iron film, which was then calcined. 50 nm diameter bare gold particles and 50 nm gold-core particles covered with a thin silica-shell were used.

Photocurrent measurements were carried out in a three-electrode photoelectrochemical cell. The working electrode consisted of the various studied photoelectrodes. A platinum mesh was used as the counter electrode, and a saturated calomel electrode was used as the reference electrode. Currents under illumination and in the dark were measured under an applied potential of 1.41 V vs. the reversible hydrogen electrode (RHE). All samples were measured in an aqueous solution of ultra-pure water (18 MΩ cm) containing 0.2 M sodium acetate. The photocurrent action spectra were obtained using a supercontinuum white-light source coupled to an acousto-optical tunable filter with a transmission bandwidth of 20 nm. The illumination spot size on the sample was ~ 1 mm in diameter and powers of 1-3 mW were used depending on the wavelength. The intensity was thus around 1-5 kW/m$^2$. The monochromated light was manually chopped and the photocurrent signal was measured using a potentiostat. An equilibration time of ~ 60 seconds after each chop allowed the current to reach a steady state value which was then measured. Photocurrent enhancements were calculated as the ratio of photocurrents recorded on areas with and without nanoparticles on the same sample. Photocurrent measurement errors were on the order of 0.1 nA. The relative error of the photocurrent enhancements was calculated by adding the relative errors of the photocurrent measurements with and without nanoparticles in quadrature. Full-field electromagnetic simulations were performed using a 2D finite difference frequency domain method (COMSOL), as shown in Fig. 4a. A single gold particle was placed either on top (20 nm embedded, to be precise) or at the bottom of an iron oxide film, followed by a TCO and a semi-infinite silica substrate. Normal-incidence transverse magnetic (TM) polarized illumination was used. The dissipated power in a small probe region (20 nm vertical x 100 nm horizontal) was calculated, both with and without the nanoparticle present. The vertical size of the probe region was chosen to take into account the distance over which carrier transport to the liquid interface is expected to be effective. Horizontally, the probe region is large enough to capture most near-field effects, but small enough to simulate the fairly high areal nanoparticle coverages used in some of our experiments. The absorption enhancement is calculated as the ratio of the dissipated power in the probe region with/without nanoparticle. We also calculated the dissipated power in the nanoparticle in order to assess plasmonic effects. In simulations employing mixed $Fe_2O_3/SiO_2$ core-shell nanoparticles on the top of iron oxide the probe region includes the shells. Absorption enhancements were area-normalized to take into account the different probe region areas in the structures with/without nanoparticles. To simulate samples on the rough, 940 nm thick, FTO substrates (experimentally 33 nm RMS, 200 nm peak-peak roughness, see Supplemental Information) an incoherent average was taken over the enhancement results obtained in simulations with 890 - 940 nm thick FTO. This incoherent average is used to account for the thickness variations in the rough FTO film, primarily smearing out interference effects. As a control experiment, we have verified that coating bare transparent conductive electrode substrates with 50 nm gold nanoparticles (both with and without silica shells) of comparable areal density as on iron oxide does not produce significant photocurrents (<1 - 2 nA at all wavelengths of interest, compared to ~ 100 nA at 590 nm in our iron oxide photoelectrodes). Therefore, the observed enhancements are true enhancements of photocurrents generated in iron oxide[35].

**References**
1. Lewis, N. S., Nocera, D. G. Powering the planet: Chemical challenges in solar energy utilization. PNAS 103, 15729–15735 (2006).
2. Chen, X., Shen, S., Guo, L., Mao, S. S. Semiconductor-based Photocatalytic Hydrogen Generation. Chem. Rev. 110, 6503 (2010).
3. Zou, Z., Ye, J., Sayama, K., Arakawa, H. Direct splitting of water under visible light irradiation with an oxide semiconductor photocatalyst. Nature 414, 625 (2001).
4. Boettcher, S. W., Spurgeon, J. M., Putnam, M. C., Warren, E. L., Turner-Evans, D. B., Kelzenberg, M. D., Maiolo, J. R., Atwater, H. A., Lewis, N. S. Energy-Conversion Properties of Vapor-Liquid-Solid-Grown Silicon Wire-Array Photocathodes. Science 327, 185 (2010).
5. Beermann, N., Vayssieres, L., Lindquist, S.E., Hagfeldt, A. Photoelectrochemical studies of oriented nanorod thin films of hematite. J. Electrochem. Soc. 147, 2456-2461 (2000).
6. Goodey, A. P., Eichfeld, S. M., Lew, K.-K., Redwing, J. M., Mallouk, T. E. Silicon Nanowire Array Photoelectrochemical Cells. J. AM. CHEM. SOC. 129, 12344-12345 (2007).
7. Kay, A., Cesar, I., Grätzel, M. New Benchmark for Water Photooxidation by Nanostructured alpha-Fe2O3 Films. J. AM. CHEM. SOC. 128, 15714-15721 (2006).
8. Atwater, H. A., Polman, A. Plasmonics for improved photovoltaic devices. Nature Materials 9, 205 (2010).
9. Yu, E. T. in Nanotechnology for Photovoltaics (ed. Tsakalakos, L.) (CRC Press, 2010).
10. Pillai, S., Catchpole, K. R., Trupke, T., Green, M. A. Surface plasmon enhanced silicon solar cells. J. Appl. Phys. 101, 093105 (2007).
11. Pala, R. A., White, J., Barnard, E., Liu, J., Brongersma, M. L. Design of Plasmonic Thin-Film Solar Cells with Broadband Absorption Enhancements. Adv. Mat. 21, 1-6 (2009).
12. Hägglund, C., Zäch, M., Kasemo, B. Enhanced charge carrier generation in dye sensitized solar cells by nanoparticle plasmons. APPLIED PHYSICS LETTERS 92, 013113 (2008).
13. Fahr, S., Rockstuhl, C., Lederer, F. Metallic nanoparticles as intermediate reflectors in tandem solar cells. Appl. Phys. Lett. 95, 121105 (2009).
14. Ding, I.-K., Zhu, J., Cai, W., Moon, S.-J., Cai, N., Wang, P., Zakeeruddin, S. M., Grätzel, M., Brongersma, M.L., Cui, Y., McGehee, M.D. Plasmonic Dye-Sensitized Solar Cells. Advanced Energy Materials 1, 52–57 (2011).
15. Bharadwaj, P., Deutsch, B., Novotny, L. Optical Antennas. Advances in Optics and Photonics 1, 438 (2009).
16. Schuller, J. A., Barnard, E.S., Cai, W., Jun, Y.C., White, J.S., Brongersma, M.L. . Plasmonics for extreme light concentration and manipulation. NATURE MATERIALS 9, 193-204 (2010).
17. Kennedy, J. H., Frese, K. W. Photooxidation of Water at alpha-Fe2O3 Electrodes. J. Electrochem. Soc 125, 709 (1978).

18. Dare-Edwards, M. P., Goodenough, J. B., Hamnett, A., Trevellick, P. R. Electrochemistry and photoelectrochemistry of iron(III) oxide. J. Chem. Soc., Faraday Trans. I 79, 2027 (1983).
19. Grätzel, M. Photoelectrochemical cells. Nature 338, 414 (2001).
20. Miller, E. L., Rocheleau, R.E., Khan, S. A hybrid multijunction photoelectrode for hydrogen production fabricated with amorphous silicon/germanium and iron oxide thin films. International Journal of Hydrogen Energy 29, 907 – 914 (2004).
21. Mavrikakis, M., Stoltze, P., Nørskov, J. K. Making gold less noble. Catalysis Letters 64, 101–106 (2000).
22. Subramanian, V., Wolf, E., Kamat, P. V. Semiconductor-Metal Composite Nanostructures. To What Extent Do Metal Nanoparticles Improve the Photocatalytic Activity of $TiO_2$ Films? J. Phys. Chem. B 105, 11439-11446 (2001).
23. Watanabe, A., Kozuka, H. Photoanodic Properties of Sol-Gel-Derived $Fe_2O_3$ Thin Films Containing Dispersed Gold and Silver Particles. J. Phys. Chem. B 107, 12713-12720 (2003).
24. Thimsen, E., Le Formal, F., Grätzel, M., Warren, S. C. Influence of Plasmonic Au Nanoparticles on the Photoactivity of $Fe_2O_3$ Electrodes for Water Splitting. Nano Lett., DOI: 10.1021/nl1022354 (2010).
25. Kelly, K. L., Coronado, E., Zhao, L. L, Schatz, G. C. The Optical Properties of Metal Nanoparticles: The Influence of Size, Shape, and Dielectric Environment. J. Phys. Chem. B 107, 668 (2003).
26. Ung , T., Liz-Marzan, L. M., Mulvaney, P. Gold nanoparticle thin films. Colloids and Surfaces A: Physicochemical and Engineering Aspects 202, 119–126 (2002).
27. Cesar, I., Kay, A., Gonzalez Martinez, J. A., Grätzel, M. . Translucent Thin Film $Fe_2O_3$ Photoanodes for Efficient Water Splitting by Sunlight: Nanostructure-Directing Effect of Si-Doping. J. AM. CHEM. SOC. 128, 4582-4583 (2006).
28. Liang, Y., Enache, C. S., van de Krol, R. Photoelectrochemical Characterization of Sprayed α-$Fe_2O_3$ Thin Films: Influence of Si Doping and $SnO_2$ Interfacial Layer. International Journal of Photoenergy, doi:10.1155/2008/739864 (2008).
29. Jensen, T. R., Schatz, G. C., Van Duyne, R. P. Nanosphere Lithography:  Surface Plasmon Resonance Spectrum of a Periodic Array of Silver Nanoparticles by Ultraviolet−Visible Extinction Spectroscopy and Electrodynamic Modeling. J. Phys. Chem. B 103, 2394-2401 (1999).
30. Liu, Z., Wang, H., Li, H.,. Red shift of plasmon resonance frequency due to the interacting Ag nanoparticles embedded in single crystal $SiO_2$ by implantation. Appl. Phys. Lett. 72, 1823 (1998).
31. Gautier, C., Cunningham, A., Si-Ahmed, L., Robert, G., Bürgi, T. Pigments based on silica-coated gold nanorods: Synthesis, colouring strength, functionalisation, extrusion, thermal stability and colour evolution. Gold Bulletin 43, 94 (2010).
32. Zhang, K., Wang, W.,  Cheng, W., Xing, X., Mo, G., Cai, Q., Chen, Z. , Wu, Z. Temperature-Induced Interfacial Change in Au@$SiO_2$ Core-Shell Nanoparticles Detected by Extended X-ray Absorption Fine Structure. J. Phys. Chem. C 114, 41-49 (2010).

33. Forman, A. J., Park, J.-N., Tang, W., Hu, Y.-S., Stucky, G. D., McFarland, E. W. Silica-Encapsulated Pd Nanoparticles as a Regenerable and Sintering-Resistant Catalyst. ChemCatChem 2, 1318 – 1324 (2010).
34. Kwon, S., Fan, M., Cooper, A. T., Yang, H. Photocatalytic Applications of Micro- and Nano-$TiO_2$ in Environmental Engineering. Critical Reviews in Environmental Science and Technology 38, 197-226 (2008).
35. Westphalen, M., Kreibig, D., Rostalski, J., Lueth, H., Meissner, D. . Metal cluster enhanced organic solar cells. Solar Energy Materials & Solar Cells 61, 97-105 (2000).

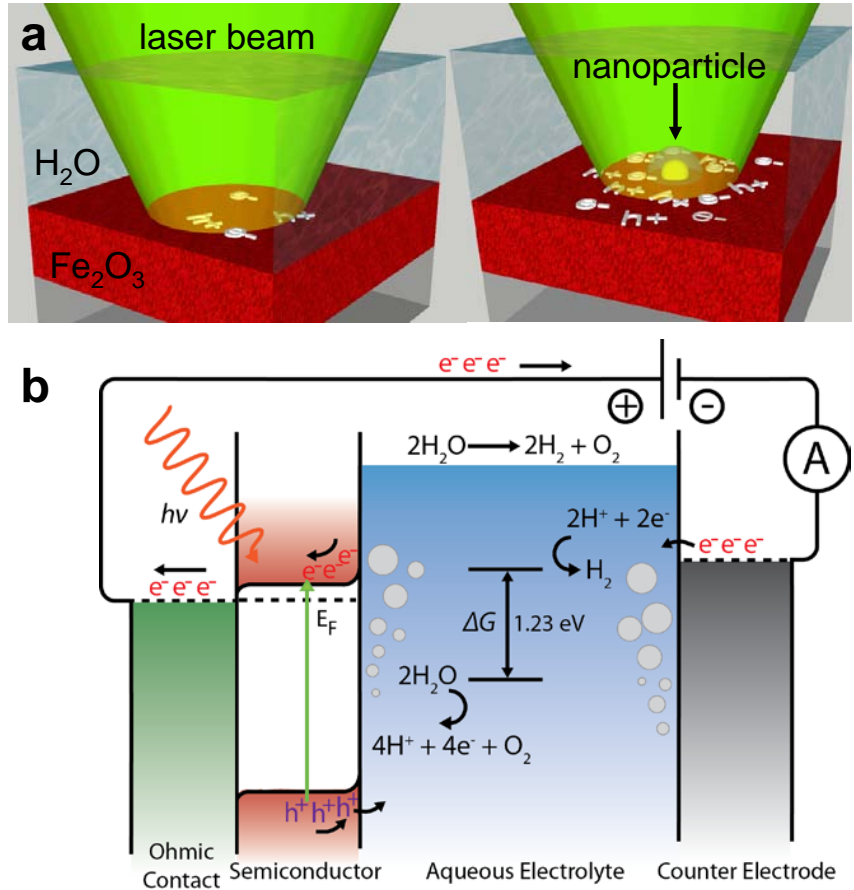

**Figure 1 | Beneficial effects of plasmonics on the performance of photoelectrochemical cells.**
**a,** A schematic that illustrates how a silica coated metal nanoparticle can effectively concentrate light near the semiconductor/liquid interface. This effect increases the number of photogenerated carriers that can reach the interface and participate in desired reactions, for example those required in water splitting to produce $O_2$ and $H_2$. **b,** Schematic band diagram depicting the semiconductor absorber (left) in contact with the liquid, where the photogenerated holes drive oxygen evolution. It is this reaction that we seek to enhance using plasmonic effects with $\alpha$-$Fe_2O_3$ as the semiconductor electrode. The generated electrons are transported to a counter electrode (right) where they can react to produce $H_2$ and close the circuit. Measurements of the photogenerated current as a function of the illumination wavelength can provide valuable clues on the potential importance of plasmonic effects.

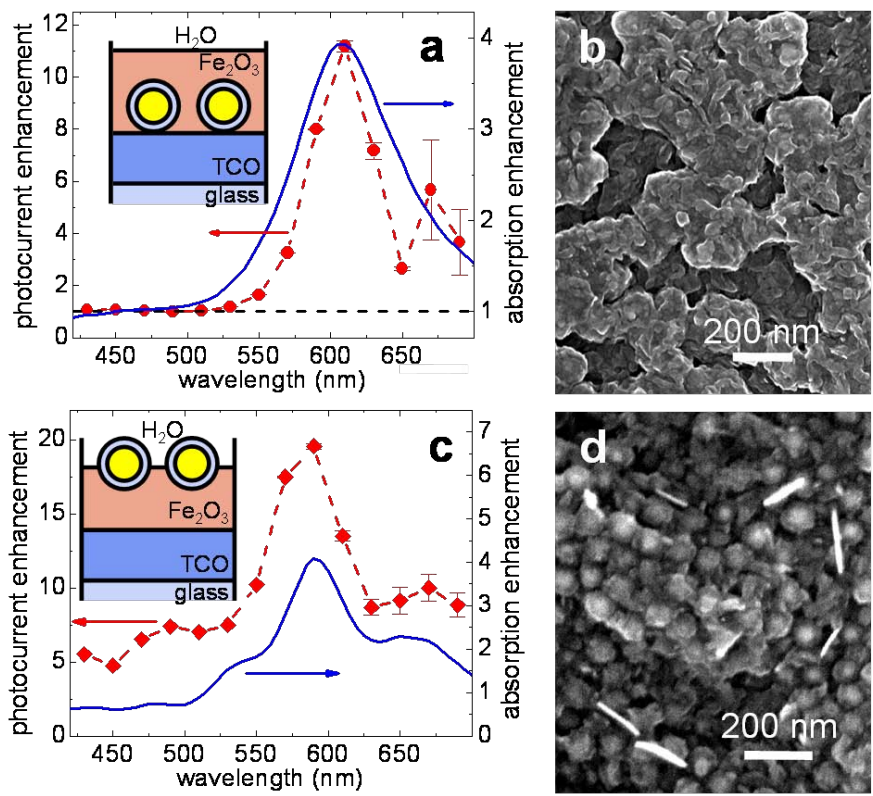

**Figure 2 | Photocurrent enhancement spectra for Au nanoparticles with a silica shell. a,c** Measured (symbols) and simulated (solid blue lines) photocurrent enhancement spectra that show the beneficial effects of placing silica-coated Au particles at the bottom/on top of a 100 nm thin $Fe_2O_3$ photoelectrode layer. The samples are shown schematically in the insets, and in scanning electron microscopy images (**b,d**). Both samples exhibit strong (> 10 x) enhancement over a relatively broad wavelength range. Electromagnetic simulations (blue lines) are consistent with a plasmonic origin of the observed enhancements in the 550 nm - 650 nm wavelength range. The sample with the particles at the $H_2O/Fe_2O_3$ interface shows an additional, more-or-less frequency-independent enhancement (about 5-8 x) that cannot be explained by electromagnetic effects, but has a chemical/physical origin (see main text).

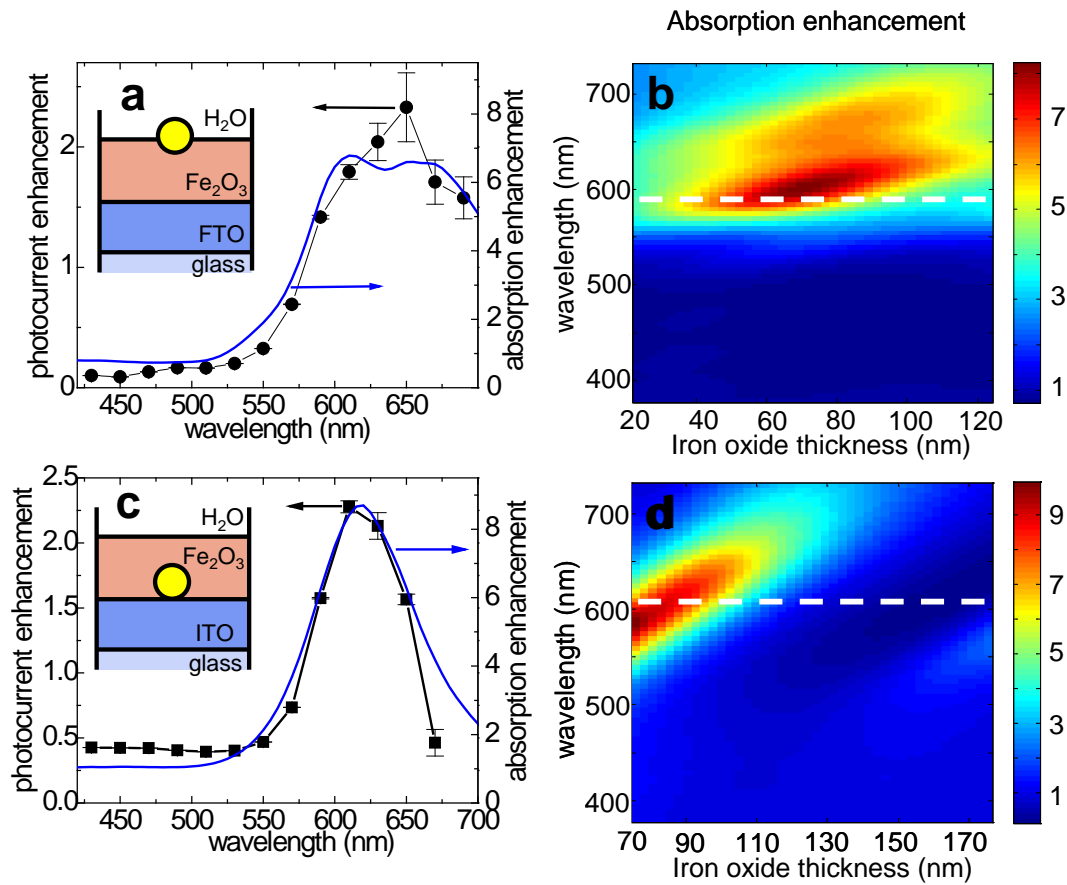

**Figure 3 | Plasmonic effects in the photocurrent enhancement spectra obtained with bare Au nanoparticles. a,c,** Measured photocurrent enhancement spectra (black symbols) exhibit one dominant spectral feature, and are well-explained by plasmonic effects (electromagnetic simulations, blue lines). In contrast to core-shell particles, bare gold particles often show reduced photocurrents compared to simulation results, possibly due to undesired charge recombination. **b,d,** Full-field electromagnetic simulations of the plasmon-enhanced absorption in a probe region near the particle and at the $H_2O/Fe_2O_3$ interface (see Fig. 4a) predict strong enhancements near the surface plasmon-resonance. The peak enhancement can be wavelength-tuned by varying the $Fe_2O_3$ thickness. This suggests that the peaks arise from an interplay between multilayer interferences and surface plasmon effects. For reference, the dashed white lines in **b** and **d** indicate the surface plasmon resonance wavelength for a 50 nm Au nanoparticle located on top of or embedded in a semi-infinite $Fe_2O_3$ respectively. It is important to note that the samples with Au particles on top of the $Fe_2O_3$ produce asymmetric photocurrent enhancement spectra, whereas the samples with particles at the bottom of the $Fe_2O_3$ film produce symmetric peaks. This observed behavior is a signature of plasmonic effects.

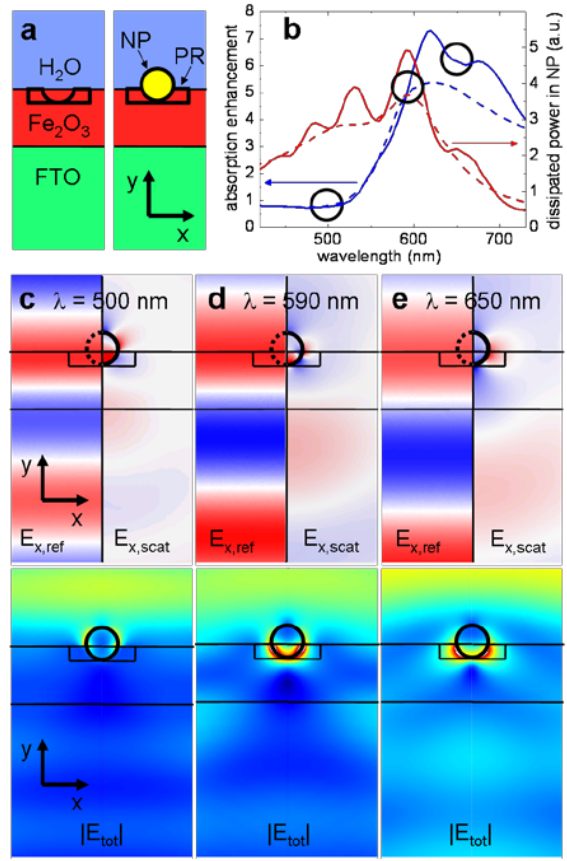

**Figure 4 | Full-field electromagnetic simulations explaining the physical origin of the photocurrent enhancements and their spectral distribution. a,** Simulation geometry with a Au nanoparticle (NP) on top of an $Fe_2O_3$ film. The probe region (PR) is the region in which absorption enhancements were calculated. **b,** The solid lines show simulations appropriate to the multilayer geometry depicted in 4a. The red line shows the absorption in the metal particle itself and the blue line shows the particle-induced absorption-enhancement in the $Fe_2O_3$ probe region. The dashed red/blue curves show the same quantities for a semi-infinite $Fe_2O_3$. In these spectra the multilayer interference effects (wavelength-dependent oscillations in the photocurrent) are suppressed, and the contributions from plasmonic effects are most easily recognized. The red curves are peaked near the surface plasmon resonance wavelength of $\lambda_{SP} = 590$ nm, where incident light is most efficiently converted to heat in the particle. The blue spectrum is strongly asymmetric, which is a signature of a plasmonic effect that is explained in Figs. c-e. **c-e,** Simulation of the E-fields in the multilayer geometry for wavelengths of 500, 590 and 650 nm, respectively. Shown are the x-component of the E-field without a NP, the x-component of the scattered field near the NP, and the modulus of the total field in the presence of the NP. For wavelengths shorter than $\lambda_{SP}$, the scattered field is out of phase with the field from the incident wave, resulting in destructive interference and low fields below the particle (**c**). For wavelengths longer than $\lambda_{SP}$, interference in the forward direction is constructive, resulting in an enhanced electric field (i.e. light intensity) in the probe region (**e**).

Supplementary Information

# Plasmon Enhanced Solar Energy Conversion

I. Thomann[1]*, B.A. Pinaud[2], Z. Chen[2], T.F. Jaramillo[2], B.M. Clemens[1], M.L. Brongersma[1]*

[1]*Geballe Laboratory for Advanced Materials, 476 Lomita Mall, Stanford, California 94305*

[2]*Department of Chemical Engineering, Stanford University, Stanford, California 94305*

* To whom correspondence should be addressed.

E-mails: ithomann@stanford.edu , brongersma@stanford.edu

The supplementary information contains nine figures in support of the main body of the text.

It discusses the following specific topics:

1. Additional information regarding the electromagnetic simulations (Figure S1)
2. Additional data supporting the conclusions of the main text (Figures S2- S4).
3. Further details on the photoelectrochemical characterization of iron oxide electrodes (Figures S7, S8).
4. Materials characterization of the iron oxide photoelectrodes (Figures S5, S6, S9)

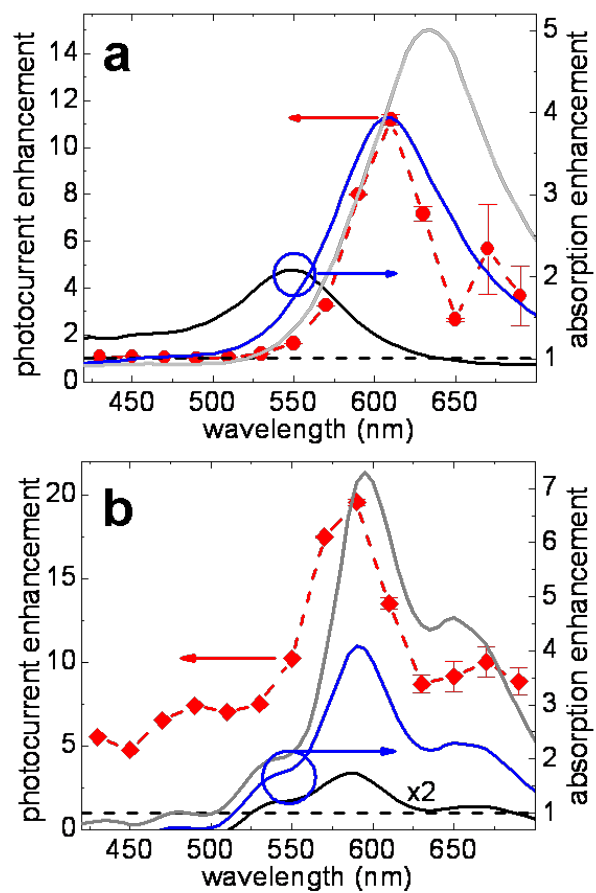

**Figure S1.** Full-field simulations employing gold nanoparticles covered with pure silica shells (black lines) cannot reproduce the experimentally observed strong enhancement peaks around 600 nm (red symbols), both for particles (a) at the bottom and (b) on top of iron oxide. The experimentally observed peak positions in Fig. 2a,c (610 nm and 590 nm, respectively) lie in between the peak positions expected from Au covered by $SiO_2$ shells (550 nm) and Au covered by $Fe_2O_3$ shells (630 nm). To capture possible imperfections in the samples, we employed two strategies: (1) simulations of a mixed $SiO_2$ - $Fe_2O_3$ shell were performed; (2) similar results (not shown) were obtained in simulations using thinner silica shells rather than mixed shells. Best agreement (apart from a wavelength-independent offset in **b**, discussed in the main text) is found with a shell consisting of an effective medium of 50% iron oxide and 50% silica (blue lines). Grey: simulations using pure iron oxide shell. Note the 2 x scaling of the black line in (b).

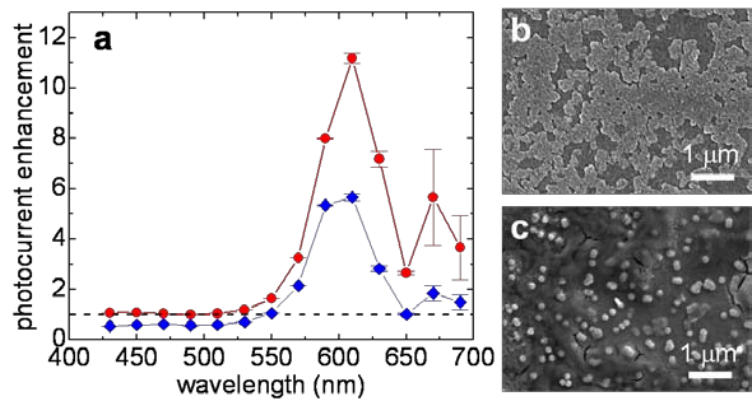

**Figure S2.** Comparison of experimentally measured photocurrent enhancement spectra in samples with low and high coverage of silica-shell gold-core nanoparticles at the bottom of the iron oxide film. (a,b) A high areal coverage (~ 35 %) of gold-core silica-shell nanoparticles (red circles) produce a peak enhancement of ~ 11 x. (a,c) A lower areal coverage (~ 5 %) of the same nanoparticles (blue diamonds) shows a reduced peak enhancement of about 5 x. The almost identical spectral shape of both curves shows that even for a large areal coverage, the silica shells prevent significant red-shifts of the spectra, which are known to result from near-field interactions between closely-spaced metal nanoparticles. Both datasets are taken on 100 nm iron oxide films. (b,c) corresponding SEM images.

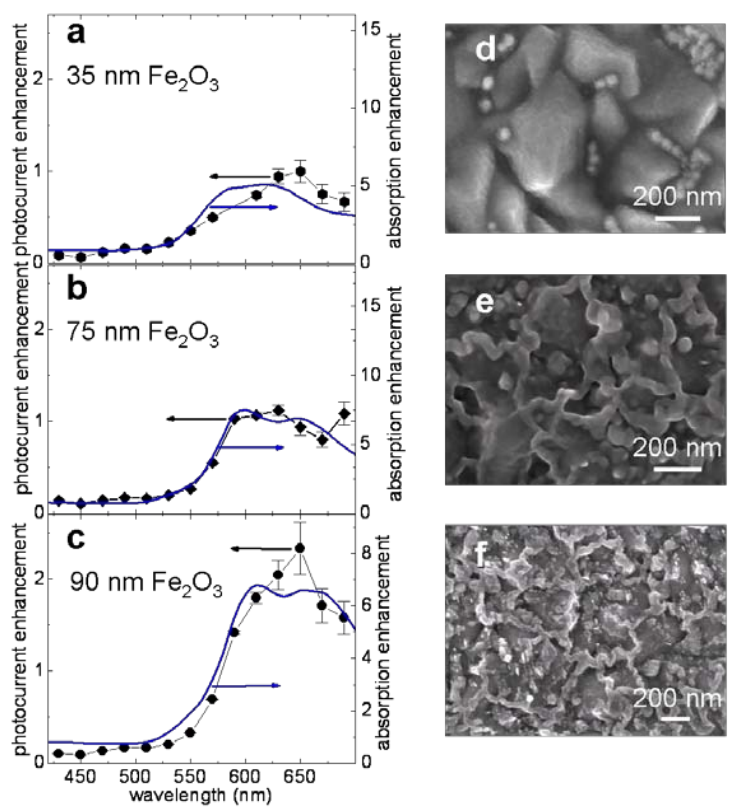

**Figure S3.** Comparison of the lineshapes of plasmonic photocurrent enhancements for bare gold nanoparticles deposited on top of different iron oxide film thicknesses. (a,d) 35 nm $Fe_2O_3$, (b,e) 75 nm $Fe_2O_3$, (c,f) 90 nm $Fe_2O_3$. (a,b,c) show measured photocurrent enhancements (black symbols) and simulated absorption enhancements (blue lines). Overall, the measured photocurrent enhancements are significantly lower than simulated ones, possibly due to recombination at trap states at the metal/semiconductor interface. (d,e,f) show the corresponding SEM images. All spectra show asymmetric lineshapes, with photocurrent suppression at short wavelengths, and with enhancements rising at long wavelengths. The asymmetry can be understood in terms of the forward-backward interference asymmetry described in the main text and appears to persist in rough films (also see Fig. S4 and S5).

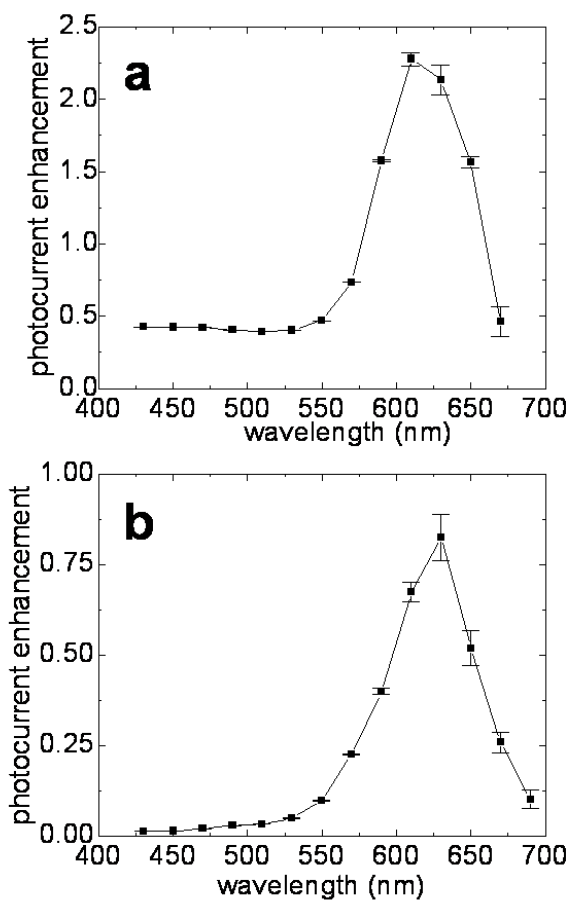

**Figure S4.** Influence of substrate roughness on photocurrent enhancements. 50 nm Au nanoparticles were deposited on top of (a) a smooth ~150 nm thick ITO substrate (0.8 nm RMS, 7 nm peak-peak surface roughness), and (b) a rough ~940 nm thick FTO substrate (33 nm RMS, about 200 nm peak-peak surface roughness). Both substrates and nanoparticles were then covered with $Fe_2O_3$ films of similar thickness. The basic spectral shape (peak position and bandwidth) of the measured photocurrent enhancements appears similar in both cases.

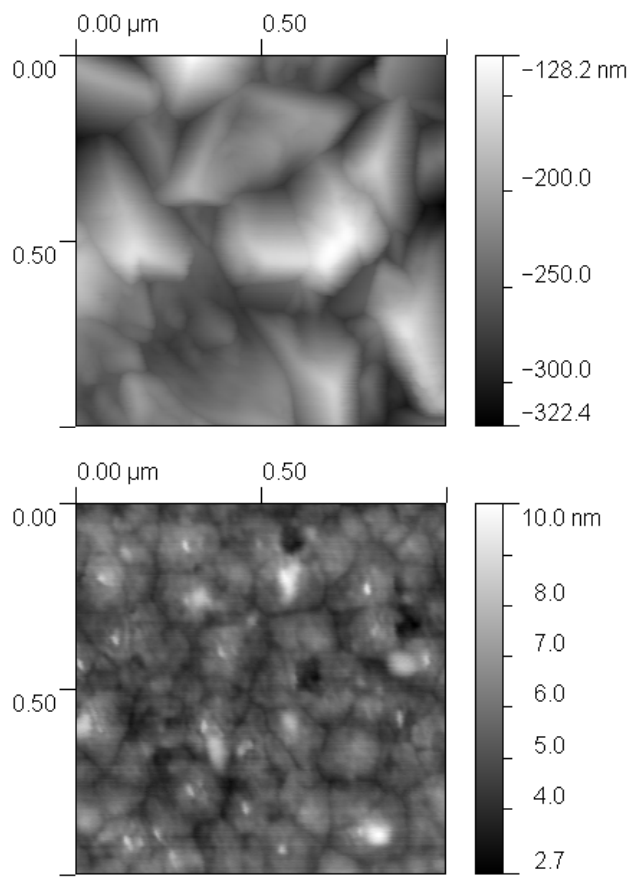

**Figure S5.** AFM images of the two types of substrates used in this work, (a) rough ~ 940 nm thick FTO and (b) smooth ~ 150 nm ITO. The measured surface roughness for the two substrates was (a) 33 nm rms, ~ 200 nm peak-peak, and (b) 0.8 nm rms, 7 nm peak-peak. Typical lateral feature sizes are ~ 300 nm in (a) and ~ 100 nm in (b).

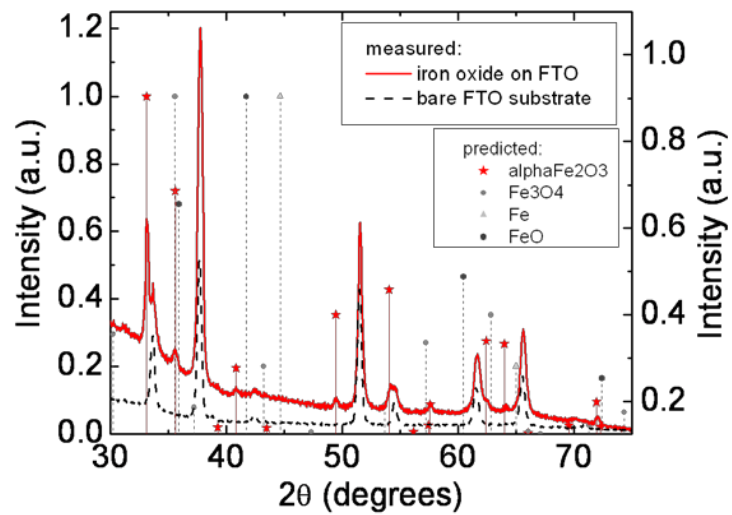

**Figure S6.** X-ray diffraction analysis of a bare FTO substrate (black dashed line) and the iron oxide films grown on this substrate (red solid line) demonstrate that the hematite ($\alpha$-$Fe_2O_3$) phase is grown. Predicted peaks from the various materials are shown as a stick plot. The alternative $Fe_3O_4$ and FeO phases, or elemental iron were not observed.

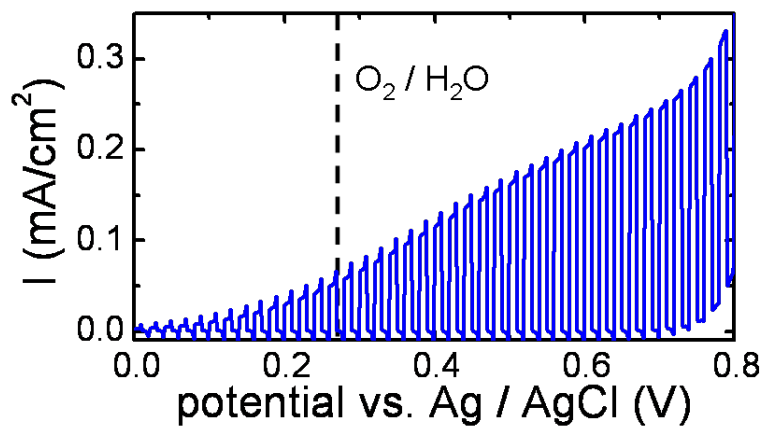

**Figure S7.** Chopped light potential sweep for a 20 nm iron oxide film, measured in a three-electrode photoelectrochemical cell setup, under ~ 1 sun illumination. The illumination beam was chopped with a 1 s period, while the potential was swept from anodic to cathodic values at 10 mV/s. The electrolyte was 0.1 M NaOH. $O_2/H_2O$: thermodynamic potential for oxygen evolution 0.266 V. The reference electrode was a Ag/AgCl electrode.

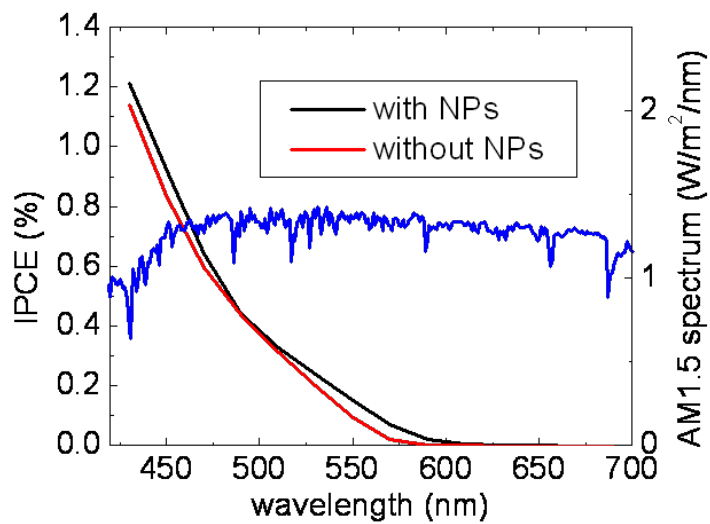

**Figure S8.** IPCE (incident photon-to-current conversion efficiency) calculated from the measured photocurrents for the structure in Fig. 2a. Black (red): IPCE with (without) nanoparticles, respectively. Blue: AM 1.5 solar illumination for reference. The AM1.5 spectrum can be multiplied with the measured IPCE, yielding an expected photocurrent spectrum. The overall, wavelength-integrated photocurrent enhancement is 1.12 x, i.e. a 12% increase over the reference structure. A 7% increase arises in the region > 490 nm, in which plasmonic enhancements are expected.

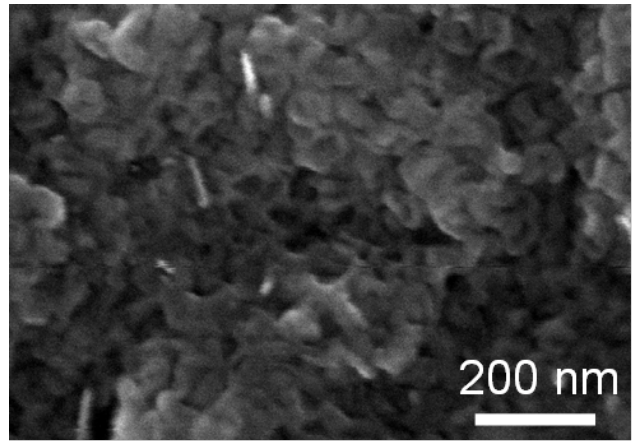

**Figure S9.** Scanning electron microscopy (SEM) image of a bare iron oxide film, corresponding to the reference region for the sample described in Figure 2d.